\documentclass[aps,showpacs,twocolumn,superscriptaddress]{revtex4}
\usepackage{graphicx}
\usepackage{dcolumn}
\usepackage{amsmath}

\begin{document}

\title{A novel active quenching circuit for single photon detection with Geiger mode avalanche photodiodes}

\begin{abstract}
\rule{0ex}{3ex}
\noindent
{\bf
In this paper we present a novel construction of an active quenching circuit intended for single photon detection. For purpose of evaluation, we have combined this circuit with a standard avalanche photodiode C30902S to form a single photon detector. A series of measurements, presented here, show that this single photon detector has a dead time of less than 40ns, maximum random counting frequency of over 14MHz, low after pulsing, detection efficiency of over 20\% and a good noise performance. This simple and robust active quenching circuit can be built from of-the-shelf electronic components and needs no complicated adjustments. 
 }
\end{abstract}

\author{M. Stip\v cevi\' c}
\email{Mario.Stipcevic@irb.hr}
\affiliation{\footnotesize Rudjer Bo\v{s}kovi\'{c} Institute,
         Bijeni\v cka 54, P.O.B. 180, HR-10002 Zagreb, Croatia}

\maketitle

\section*{Introduction}

Experiments in quantum information and communication have concentrated on studying 
photons, which among all elementary particles, have unique properties of being easily produced in abundance, easily manipulated, easily transmitted to large distances without interaction with the surrounding matter and yet relatively easy to detect. Main tools in research an application of  quantum information and communication, namely fiber optic light guides and new ultra-bright sources of entangled photons based on parametric downconversion in nonlinear crystals operate in the near infrared range of 700-1550nm \cite{downqkd}. 
\\

Photomultipliers, as traditional photon detectors, have very small quantum efficiency and a large noise in that wavelength range. The next most mature technology of detecting single photons is based upon avalanche photo diodes intended for single photon detection, so called SPAD's (Single Photon Avalanche Diode). In principle, silicon and GaAS SPAD's offer excellent quantum efficiency, high gain, mechanical robustness, possibility of miniaturization, low power consumption and relatively low cost.
However, there are only a few commercially available SPAD-based photon detector modules for visible and near IR range on the market. Most of them are optimized for fast timing rather than for a good detection efficiency in the NIR range. Furthermore, prices of these modules are quite high when comparing to the price of the main component, namely the avalanche photodiode.  
\\

In this paper, we present a single photon detector (i.e. photon counter) based upon a novel construction of an active quenching circuit. The detector makes use of the popular avalanche photodiode C30902S which has a peak detection efficiency at 830nm, and is quite sensitive in the range 400-1000nm. The active quenching circuit can be built from of-the-shelf electronics components and needs no complicated adjustments. 

\section*{Passive versus active quenching}

Detection of randomly arriving photons with SPAD's requires either passive or active quenching techniques.
\\

There are three principal sources of avalanches in a SPAD: 1) an incoming photon, which converts into a carrier; 2) the afterpulse (a carrier leftover from a previous avalanche trapped and then released from an impurity at a latter time) and 3) the dark count (a carrier created by thermal excitation in the sensitive area of the SPAD). These mechanisms are well described elsewhere \cite{dautet}.
\\

\begin{figure}[h]
\centerline{\includegraphics[width=40 mm,angle=0]{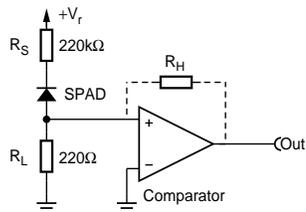}}
\caption{ A typical passive quenching circuit for single photon detection. } 
\label{passive}
\end{figure}

A typical passive quenching circuit \cite{rarity1} is shown in Fig.\ref{passive}. It consists of inversely biased SPAD in series with a large current limiting resistance ($R_S$=220k$\Omega$) and a small signal picking resistor ($R_L$=220$\Omega$). The SPAD is kept at the inverse bias voltage $V_R$ somewhat higher than the breakdown voltage $V_{BR}$. The "overvoltage" $V_{over}=V_R - V_{BR}$ defines the detection efficiency (D.E.) and noise of the SPAD. For example, the D.E. of C30902S for overvoltage in the range from 0V to 25V monotonically rises from 0\% to cca. 58\% \cite{datasheet}. In most of the tests conducted in this paper, the SPAD was operated 3.5V above the breakdown voltage yielding the D.E. of $\approx$15\% at 830nm.

If $R_S$ is chosen sufficiently large, an avalanche will cause a sudden drop 
of the bias voltage to a level below the breakdown voltage causing in turn the avalanche to cease within a sub-nanosecond time, that is to "quench". 
A condition for successful quench is that $V_{over}/R_S$ is smaller than the {\em latch current} of the given SPAD.
After the quench, voltage across the diode recovers toward its initial value following the exponential law with the time constant $\tau = R_S C_{SPAD}$, where $C_{SPAD}$ is the capacitance of the reversely polarized SPAD (plus any parasitic capacitances present in the actual circuit). During the bias recovery SPAD operates at lower overvoltages and has lower detection efficiency than nominal. Furthermore, avalanches that are occurring during the bias recovery will result in a pileup, which can "paralyze" the comparator and result in a prolonged dead time \cite{parart}.
Both effects, namely the pileup and lowering of the photon detection efficiency due to the previous avalanches, lead to appearance of {\em time correlations} between detected photons even if the incident photons were not correlated. These fake correlations, which appear in the detector itself, are unwanted in many applications such as: time resolved spectroscopy, quantum cryptography, random number generators, etc.
\\

In order to maximize efficiency and minimize fake correlations, $\tau$ should be kept as small as possible. Unfortunately
the minimum value of $\tau$ is limited by the latch current of the SPAD and the capacitance $C_{SPAD}$:

\begin{equation}
\tau \geq \frac{V_{over}}{I_{LATCH}} C_{SPAD}.
\label{fmeas}
\end{equation}

For example, for C30902S with $C_{SPAD} \approx 3$pF and $I_{LATCH} \approx 50\mu$A, the minimum value of $\tau$ at an overvoltage of 15V is about 1$\mu$s. The maximum counting frequency of such a setup is only about 0.25MHz. Furthermore, the detection frequency at the output of the detector is a highly nonlinear function of the incoming photon frequency for the detection frequency greater than a few tens of kilohertz.
\\

The solution to problems of low counting frequency, nonlinearity, long dead time and correlations is found in active quenching approach \cite{rarity2} where both quenching and bias voltage recovery are speed up by active electronics elements. In that way, except during a negligibly short transition interval, the SPAD is either completely "dead" or fully sensitive. The length of the dead time interval is thus well defined as opposed to the passive quenching. Active quenching can greatly improve the linearity, enhance the maximum attainable photon counting frequency 
and minimize fake correlations.

\section*{The active quenching circuit}

\begin{figure}[h]
\centerline{\includegraphics[width=80 mm,angle=0]{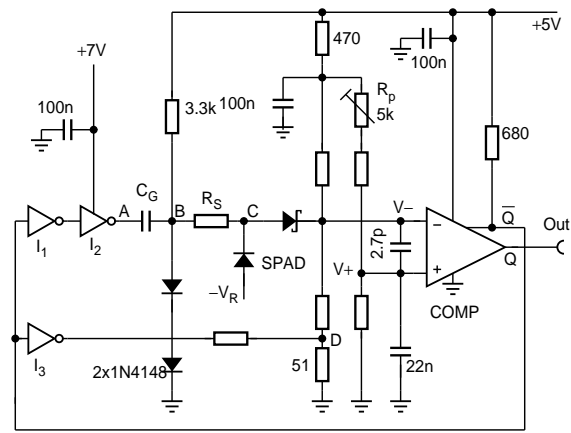}}
\caption{ The active quenching circuit. } 
\label{aqcircuit}
\end{figure}

Our active quenching circuit shown in Fig.\ref{aqcircuit} operates as a non-retriggerable monostable, triggered by the avalanche signal from the SPAD. The circuit performs three functions: amplifier of the tiny avalanche signal from the SPAD, a discriminator and a pulse shaper. A distinctive novelty of this circuit is that it uses a DC decoupling capacitor ($C_G$) to produce a negative quenching signal well below the ground without the need for a negative power supply. Furthermore, the decoupling capacitor operates at a low voltage (on the order of $V_{over}$) regardless of the high voltage required for biasing the SPAD.
The circuit works in the following way.

Let us suppose that initially the SPAD is in non-conductive state and that the output Q of the comparator AD8611 is at the low logic level (0V). A small current flowing through the resistor $R_S$ ensures that $V_-$ is slightly above the $V_+$, just enough to keep safely the output of the comparator at 0. The optimal difference $V_- - V_+$ (cca 25-30mV) can be set by potentiometer $R_p$. The goal of this setting is to have the discriminating threshold as low as possible and yet high enough to safely avoid sustained oscillations of the circuit (i.e. bistable behavior). The passive network is designed such that if the current flowing through the Schottky diode BAT83 would be absent, then the negative input of the comparator would go some 150mV below the positive input, forcing the comparator to switch to logic 1 (+3.6V).
\\

\begin{figure}[h]
\centerline{\includegraphics[width=80 mm,angle=0]{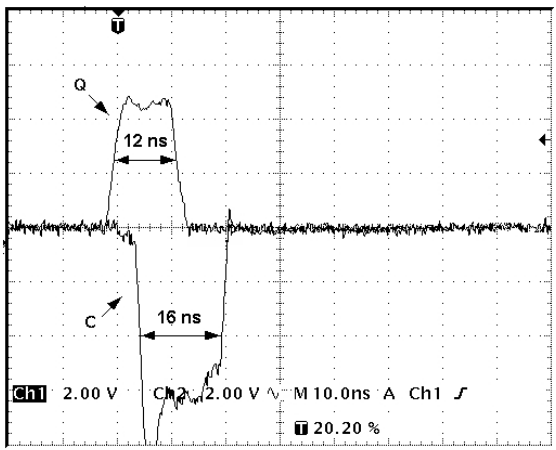}}
\caption{ Digital recording of pulses at the output of the A.Q. circuit and at the point $C$, following a detected photon. } 
\label{scope1}
\end{figure}

The detection cycle begins 
when an avalanche happens in the SPAD. The avalanche current will lower the voltage at the point C and consequently at the point $V_-$. If voltage level at $V_+$ is set appropriately, the avalanche current will cause $V_-$ to go below $V_+$ and consequently the comparator's output Q will go to logic 1, after some propagation time. 
Logic 1 at the output Q will cause change of the state of the inverter $I_2$. 
The output of the inverter $I_2$ makes a transition from $\approx$ +7V to $\approx$ 0. This voltage step of $\approx$ -7V  will be transferred (via the capacitor $C_G$ and resistor $R_S$) to the point $C$ (Fig. \ref{scope1}) and quench the avalanche. Because now, the SPAD is biased below its breakdown voltage and the Schottky diode is switched off (i.e. in non-conductive state) the detector is "dead" - it will not detect any further incoming photons. The SPAD will be held in this state for some period $t_{Q}\approx$16ns allowing carriers created during avalanche to recombine. Namely, during the quench $V_-$ is some 150mV below $V_+$ thus sustaining the quench state. However, eventually, the inverter $I_3$ will make a positive step of about 500mV at the point D thus injecting enough current for $V_-$ to go above $V_+$, which will (after propagation delay times) cause Q to rest at 0 and $I_2$ to go to logic high state and end the quench period. This completes the detection cycle after which the SPAD is again biased at its nominal operating voltage and fully ready to detect further photons. 

The bias recovery at the end of the quenching period is similar to the passive quenching where the SPAD is charged through a resistor $R_S$, except that here the biasing resistance $R_S$ is chosen so small that charging of the SPAD takes only 2-3 nanoseconds. It is only during this negligibly short time that the SPAD is operating at a voltage lower than nominal.
It is possible to achieve even quicker bias recovery (i.e. higher $dV_R/dt$) by slight modification of this circuit, however, according to our experimentation, too quick recovery of bias will result in an elevated afterpulse probability, an effect already noticed for this particular type of SPAD \cite{Cova96}. 
\\

In conclusion, each avalanche causes a well-defined logic pulse at the output of the active quenching circuit. Photons hitting the SPAD during the dead time can not produce strong avalanches and therefore are ignored and no pulse is produced.

\section*{The experimental setup}

\begin{figure}[h]
\centerline{\includegraphics[width=80 mm,angle=0]{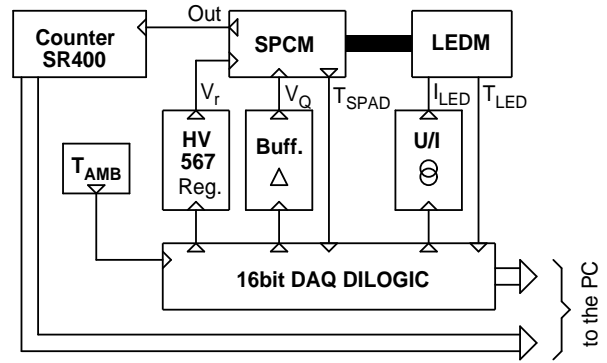}}
\caption{ Experimental setup for evaluating of the active quenching circuit. } 
\label{setup_block}
\end{figure}

The experimental setup for evaluating of the active quenching circuit is shown in Fig.\ref{setup_block}. The single photon counting module (SPCM) consists of the active quenching circuit and thermoelectrically cooled SPAD C30902S. 
The LED module described above is optically connected to the SPCM via an opaque pipe. It incorporates an independent temperature stabilizing circuitry.
All relevant parameters of the setup (SPAD's temperature and operating voltage, quenching voltage, light intensity from the LEDM, etc.) can be controlled and/or monitored by a PC computer.


The single photon detector (SPCM) has been realized using the above described active quenching (A.Q.) circuit and a popular (well known) single photon avalanche diode C30902S from PekinElmer. For this type of photodiode the optimum values of the unspecified components in the A.Q. circuit (Fig.\ref{aqcircuit}) are: $R_S$=1k, $C_G$=330pF and $R_p$ $\approx$ 2.50k. These values were optimized for lowest dead time and their values are not very critical: variations of $\pm$10\% barely have any influence on the operation of the circuit which gives us a confidence that the  circuit is robust under temperature, aging and initial tolerance component variations. The value of $R_p$ has to be trimmed according to the above described procedure because its optimal value depends on the offset of the particular comparator chip. The SPCM also incorporates a temperature stabilizing circuit described below. The desired temperature of the SPAD can be set via an external voltage supplied by the DAQ module.
\\

Because the breakdown voltage of the SPAD C30902S rises with temperature (temp. coefficient $\approx$ +0.7V/K), it was necessary to provide a temperature stabilization.
To that end, the SPAD was coupled via a piece of extruded aluminum to the thermoelectric cooling module (Peltier) with a temperature adjustable in the range -30$^o$C to +20$^o$C. The achieved stability of temperature of the SPAD's casing was better than 0.05K during any of the measurements. The temperature is measured by the electronic thermometer chip LM335.
\\

The negative bias voltage for the SPAD ($V_R$, Fig. \ref{aqcircuit}) was provided by the ORTEC 567 NIM voltage source. This instrument features an analog input that makes possible to set the output voltage in the range of 0 to -3000V. The output voltage is proportional to the DC control voltage in the range of 0 to 10V. By means of a 16 bit, DAC whose output can swing from 0 to 10V and a suitable passive voltage divider, we were able to set the bias voltage between 0 and 300V in steps of only 10mV.      
\\

\begin{figure}[h]
\centerline{\includegraphics[width=85 mm,angle=0]{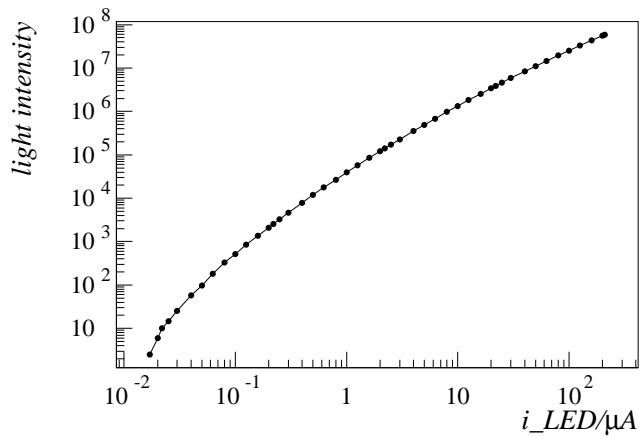}}
\caption{ Light vs. current characteristic of temperature stabilized LED module. Measurement points (dots) and the fitted polynomial curve are shown. } 
\label{cled}
\end{figure}

A specially calibrated Light Emitting Diode Module (LEDM) was used to perform linearity measurements of the SPC. LEDM is basically a temperature stabilized LED calibrated in such a way that we know quite precisely number of emitted photons (up to a multiplicative constant) as a function of the current supplied to the diode. 
We have used Hamamatsu LED type L7868, a high efficiency emitter at $\approx$ 670nm mounted in a metal casing with a flat glass window. The LED is coupled to a Peltier element in much the same way as the SPAD and held at a constant temperature of (17.50$\pm$0.05)$^o$C  in order to minimize light intensity variations with changes of the operating current and/or the room temperature. Namely, the light intensity of this LED has a temperature coefficient of -0.58\%/K.
Calibration was done by measuring the intensity of emitted light as a function of the current flowing through the LED, in the range 0.02$\mu$A-200$\mu$A. The intensity was measured by a photomultiplier (HAMAMATSU type R636-10) for LED currents in the range of 1-200$\mu$A assuming linearity between the intensity of the light falling on the photocathode and the PM tube current. For the low end current range (0.02-1$\mu$A) we have used a fast photon counter (Electron Tubes type DM0016C) assuming linearity between the incident light intensity and the frequency of counts, after corrections for the dark counts and the dead time. The smooth curve representing light intensity vs. LED current(Fig. \ref{cled}) is obtained as a polynomial fit through the measurement points (round dots) shown logarithmic scale.

The whole calibration procedure and the LED module itself will be described in detail elsewhere \cite{ledm_piblication}.

The LEDM was optically coupled to the SPC module via an 8cm long, 3mm inner diameter opaque rigid pipe. The pipe was hermetically sealed at both ends to prevent frosting of both the SPAD and the LED, while its rigidness ensured stability of the optical path.
\\

The SR400 Dual channel gated photon counter from Stanford Research was used as a frequency counter and a single event detector. 

All components of the evaluation setup are connected to a PC computer (Fig. \ref{setup_block}). The photon counter is connected directly via its RS-232 link, whereas all other components are connected to a custom-made 16 bit I/O data acquisition module provided by DILOGIC. The DAQ module itself is connected to the PC through the RS-232 port. A specialized computer program written in C was used to efficiently control and monitor the whole setup thus forming a fully automated data acquisition system. 
Series of tests have been performed with this setup to characterize the active quenching circuit.
\\

In the rest of the paper we discuss different test and their results.

\section*{Temperature coefficient of the breakdown voltage}

As a first check of our measurement setup, breakdown voltage of the SPAD was measured as a function of its temperature. This function is known to be linear with a temperature coefficient of $\approx$ 0.7V/K (see C30902E/S datasheet \cite{datasheet}). The breakdown voltage $V_{BR}$ is defined as the lowest voltage at which saturated avalanches start to occur in a SPAD kept in total darkness. For the purpose of photon counting, SPAD's must be operated at voltages above the $V_{BR}$, that is in so called "Geiger mode".  \\

The breakdown voltage, $V_{BR}$, is specific for each individual SPAD and the manufacturer only guarantees that it is somewhere in the range 200-250V at 21$^o$C. Using the SR400 in the counter mode our DAQ program was able to quickly and reliably find the breakdown voltage at any temperature $T_{APD}$ by automatically adjusting the operating voltage of the SPAD such that avalanches just start to occur in a complete darkness. 

\begin{figure}[h]
\centerline{\includegraphics[width=85 mm,angle=0]{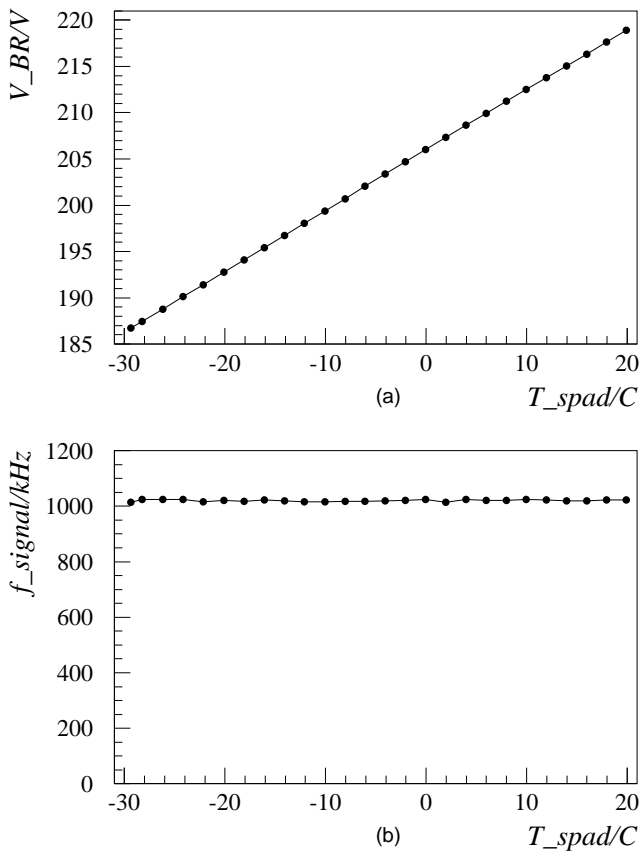}}
\caption{ (a) The breakdown voltage, $V_{BR}$, as a function of the SPAD temperature. The solid curve is a linear fit through the measurement points indicating tempco of 0.655 V/K. \\
(b) Response to a constant light at constant overvoltage, as a function of operating temperature. } 
\label{vbrvstapd}
\end{figure}

Fig. \ref{vbrvstapd}a) shows a highly linear scaling of the $V_{BR}$ with the temperature. The solid curve is a linear fit through the measured points indicating temperature coefficient of 0.655 V/K, which is in a good agreement with the value given in the datasheet. To verify that the breakdown voltage was correctly determined, at each measurement point we have also checked response of the SPAD at an overvoltage of 3.5V when illuminated by a constant light intensity from the LEDM. The intensity of light is chosen such as to give approximately 1MHz of counts (detections) at +20$^o$C. Since the detection efficiency of C30902 at 690nm is not sensitive to temperature \cite{datasheet} then if the breakdown voltage were determined correctly, one would expect that the response stays constant at any temperature, which is indeed the case, as shown in Fig. \ref{vbrvstapd}b).

\section*{Noise (dark counts)}

An important issue in photon counting technique is the noise, that is the frequency of counts in complete darkness. Of course, one would like a detector to have the noise as small as possible and detection efficiency as high as possible. However for SPADs, as will be shown, low noise operation is in contradiction with high detection efficiency and therefore an optimal solution should preferably be tailored for the desired application.\\ 

\begin{figure}[h]
\centerline{\includegraphics[width=85 mm,angle=0]{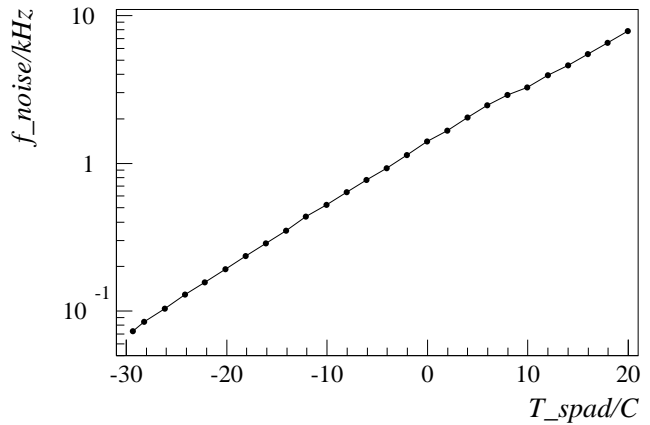}}
\caption{ Frequency of dark counts as a function of the temperature of the SPAD. } 
\label{nvstemp}
\end{figure}

Thermal excitations in a SPAD will provoke avalanches, and thus contribute to the noise. Therefore, for low noise applications it is essential to cool down the SPAD. In order to see how temperature of the SPAD affects the noise, we have measured frequency of dark counts as a function of the temperature, while keeping the overvoltage $V_{over}=V_R-V_{BR}$ fixed at 3.5V (Fig. \ref{nvstemp}). One can see that the frequency of dark counts rises nearly exponentially with the temperature of the SPAD chip, which is in accordance with thermal production of free carriers in semiconductors.

\begin{figure}[h]
\centerline{\includegraphics[width=85 mm,angle=0]{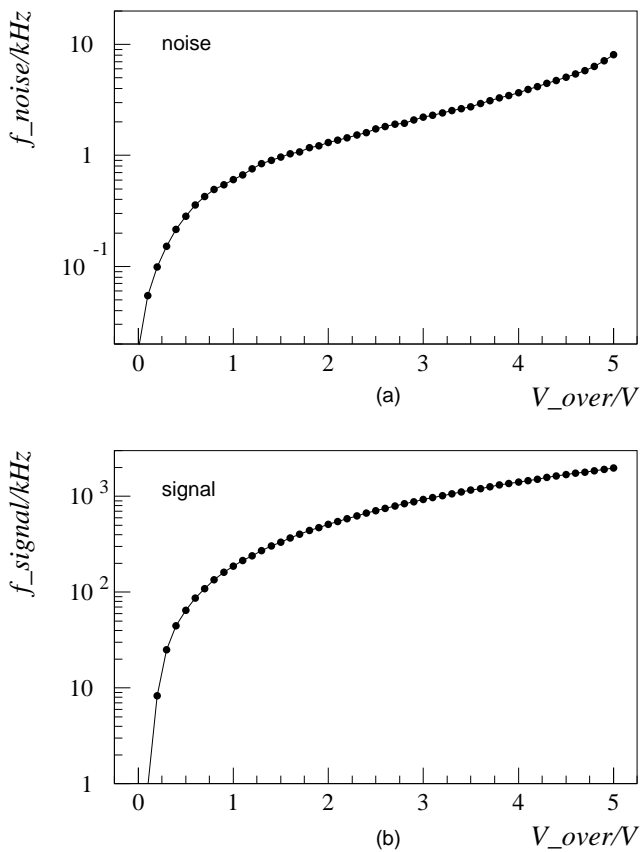}}
\caption{ The dark counts (a) and the signal (b) of the SPCM as functions of the overvoltage at a fixed temperature of the SPAD. The signal is defined as the dark counts subtracted response of the SPCM to a faint light source giving $\approx$30.000 detected photons per second. } 
\label{nsvsvplus}
\end{figure}

On the other hand, at a fixed temperature, both noise (dark counts frequency) and detection efficiency rise with overvoltage. This is illustrated by measurements of the noise (Fig. \ref{nsvsvplus}a) and signal (Fig. \ref{nsvsvplus}b) as functions of overvoltage, at a SPAD temperature of 6$^o$C. Here, "signal" is defined as the detector response to a faint light source giving approximately 30.000 detected photons per second at $V_{over}=3.5$V (noise subtracted). 
These curves are actually (a part of) the photodetection efficiency curve given in the datasheet of the SPAD.
The two curves should be alike, however in the noise curve one can see a small rise starting at about $V_{over}\approx4.5$V, which is due to the increasing inability of the A.Q. circuit to properly quench the SPAD at higher overvoltages. This effect is not seen in the signal curve because relatively rare afterpulses are overwhelmed by the signal.  

\begin{figure}[h]
\centerline{\includegraphics[width=85 mm,angle=0]{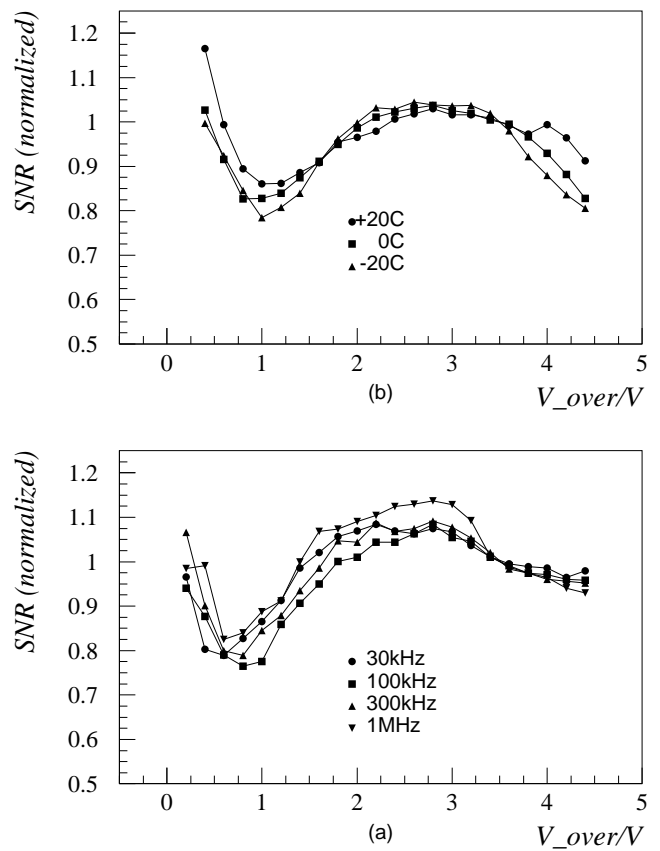}}
\caption{ (a) Signal/noise ratio of the SPCM as a function of temperature of the SPAD.\\
(b) Signal/noise ratio as a function of the detection frequency. All functions are normalized to 1 at $V_{over}=3.5$V. } 
\label{snrvsvplus}
\end{figure}

We see that both noise and the signal rise with the overvoltage. However, interestingly, the signal/noise ratio (SNR) as a function of overvoltage has a broad maximum quite independent on both temperature and incident light intensity (signal). 
Fig. \ref{snrvsvplus}a) shows SNR functions (signal of 30kHz at $V_{over}=3.5$V) at three different temperatures. 
Fig. \ref{snrvsvplus}b) shows SNR functions for signals of 30kHz,100kHz,300kHz and 1MHz at $V_{over}=3.5$V, all at a fixed temperature of $-25^o$C. 
To facilitate the comparison between them, all functions have been normalized to 1 at $V_{over}=3.5$V. 

From these measurements we conclude that for the C30902 an overall optimal SNR is achieved at an overvoltage of $\approx2-3$V where the detection efficiency is in the range of 5-15\%. Nevertheless, higher detection efficiency can be achieved at cost of higher noise or additional cooling.

\section*{Dead time}

\begin{figure}[h]
\centerline{\includegraphics[width=80 mm,angle=0]{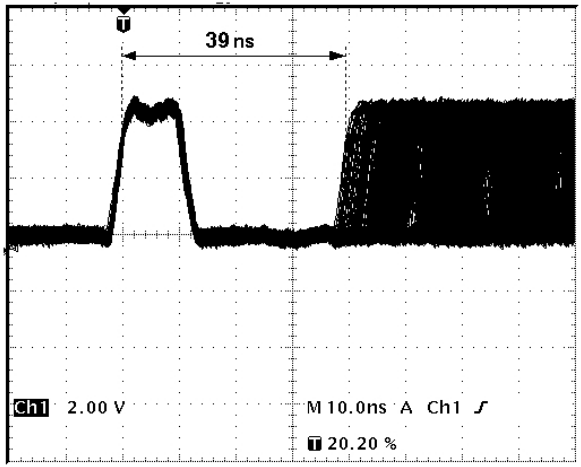}}
\caption{ Pileup of the output signal of the A.Q. circuit, corresponding to many detected photons, shown on the digital storage oscilloscope. The dead time of the photon detector, defined as the minimum time lag between the trigger and the next detections is cca 40ns. } 
\label{dtscope}
\end{figure}

The dead time of the A.Q. circuit was measured by connecting its output to the digital  storage oscilloscope (Tektronix TDS3054B) and letting many self-trigerred events to pile up one on top of the other, as shown in Fig.\ref{dtscope}. The dead time ($d$) is then determined as the minimum "distance" between a triggered event and following events. This measurement yields $d\approx 39ns$.

\section*{Linearity}

\begin{figure}[h]
\centerline{\includegraphics[width=85 mm,angle=0]{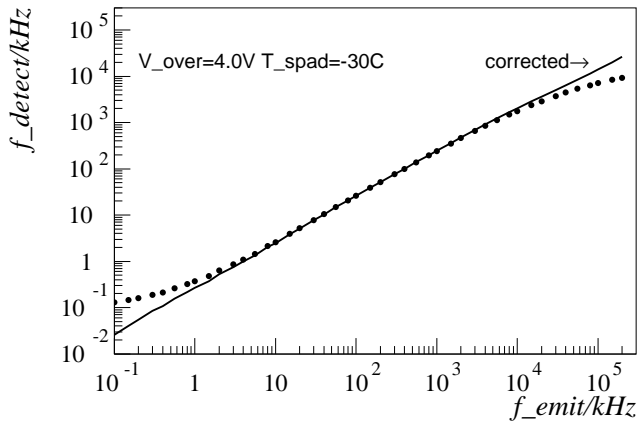}}
\caption{ Measured frequency at the output of the photon counter ($f_{meas}$) plotted against frequency of photons falling upon the SPAD ($f_{emit}$). Points present actual measurement, whereas the solid curve depicts detector's response corrected for noise and dead time. } 
\label{linplot}
\end{figure}

Closely related to the dead time is the linearity of detector's response. It can be shown (see for example \cite{hamamatsunote}) that for incident light of steady intensity (i.e. when photon arrival intervals obey exponential probability density function) the measured photon detection frequency is a monotonic function of the true detection frequency:

\begin{equation}
f_{meas} = f_{dark} + \frac{f_{true}}{1+f_{true}d}
\label{fmeas}
\end{equation}

where $f_{dark}$ is the frequency of dark counts and $d$ is the dead time. 
Fig. \ref{linplot} shows measured frequency $f_{meas}$ as a function of the frequency of the emitted photons  from the LEDM falling upon the sensitive area of the SPAD, $f_{emit}$.   

(Actually, as explained above, $f_{emit}$ is only known to a multiplicative constant and in the figure it is scaled to approximately reflect the detection efficiency of the detector.) The deviation from linear response at the low frequency end is dominated by the dark counts, whereas at high frequencies it is dominated by the dead time of the A.Q. circuit. By inverting the Eq.(\ref{fmeas}) both effects can be corrected for:

\begin{equation}
f_{corr} =  \frac{f_{meas}-f_{dark}}{1-f_{meas} d}.
\label{fcorr}
\end{equation}

The solid line in the Fig.\ref{linplot} depicts the corrected detector response which shows a very good linearity ($<$3\% deviation) up to over 10MHz.

\section*{Maximum counting frequency}

\begin{figure}[h]
\centerline{\includegraphics[width=85 mm,angle=0]{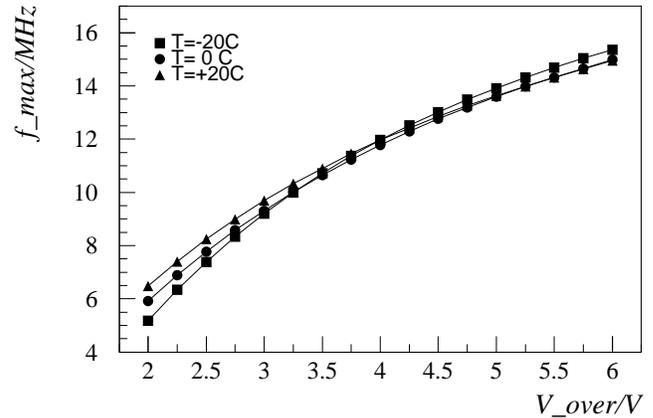}}
\caption{ The highest attainable counting rate as a function of overvoltage were made at three fixed operating temperatures of the SPAD. } 
\label{fmaxvsvplus}
\end{figure}

It is well known that for a sufficiently strong incident light a SPAD will be paralyzed (i.e. in continuous conducting state) \cite{parart}. In such a condition, no counts will appear at the output of the detector. On the other hand, in darkness the frequency of counts will be low. Obviously, somewhere in between, for some intensity of light, there will be a maximum attainable counting rate.
The maximum rate of counts is a parameter of interest for applications requiring high dynamic range of light level measurement and for optical quantum random number generators \cite{weiunfuterrng,qrbg}.
Measurements of the highest attainable counting rate as a function of overvoltage were made at three temperatures of the SPAD (-20, 0, +20) $^o$C and are summarized in Fig. \ref{fmaxvsvplus}. 

One can see that the highest attainable rate rises with the overvoltage and is virtually independent of the temperature. This can be explained by the same qualitative behaviour of the detection efficiency of the SPAD that also rises with the overvoltage and is is independent of the temperature (see \cite{datasheet}). The fact that the maximum rate being independent of the temperature is especially interesting for applications that require high counting rate and a low power consumption, such as random number generators. In that case, cooling of the SPAD (which consumes a lot of power) is not necessary. The overall maximum counting rate of the present A.Q. circuit is limited by the maximum overvoltage  at which the circuit is still capable of reliably quenching the avalanches, $V_{over}\leq$5V. At a higher overvoltage, avalanches afterpulse probability becomes enhanced.

\section*{Dynamic response}

\begin{figure}[h]
\centerline{\includegraphics[width=85 mm,angle=0]{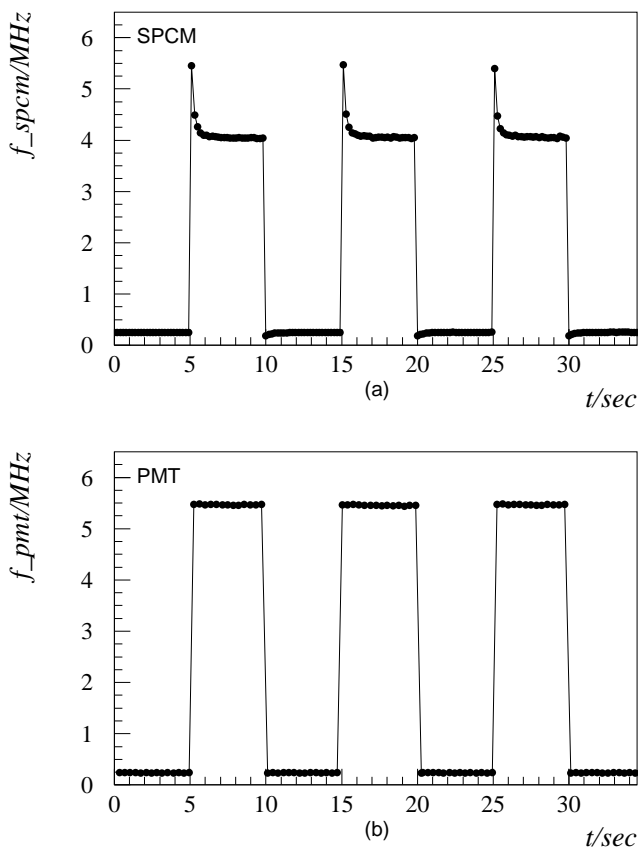}}
\caption{ Response of the SPCM photon detector module (a) and the PMT based photon counter DM0016C (b) to a pulsed light source (5 seconds low intensity, 5 seconds high intensity). } 
\label{alternate}
\end{figure}

The dynamic response of our photon detector was measured as a response to a pulse-modulated light  (Fig. \ref{alternate}a). Intensity of light coming from the LEDM was sharply switched between a low level (250kHz) and a high level (5.5MHz) at a rate of 10 seconds. The same measurement was made on the photomultiplier based photon counter DM0016C of SensTech Ltd (Fig. \ref{alternate}b). Both responses are shown corrected for respective dead times (39ns for the SPCM, 18ns for the DM0016C). While response of the DM0016C follows swiftly and precisely changes of the light intensity, the SPCM shows transitional phenomena that can be explained in terms of temperature fluctuation of the SPAD chip.
\\
  
Namely, after the jump from low to high intensity, at first moment SPAD detects expected number of photons (the same as DM0016C) but within seconds the detection frequency exponentially drops to a lower level. This can be explained by the heat building up in the SPAD's chip due to the avalanche current. The heat cannot be effectively removed away because of the thermal resistance between the chip and the metal casing (which is actually being held at constant temperature). As a consequence, at high detection rates chip operates at a temperature higher than nominal thus effectively operating at a lower overvoltage and therefore exhibiting a lower detection efficiency. 

A similar test was performed in \cite{rarity2} with a different detection cicuit
which supports the thesis that the observed transitional fluctuations in our SPCM are due to the SPAD itself and are not an artifact of the A.Q. circuit.
The transitional phenomena apparent at high counting frequencies do create a long-range time correlations between detected pulses which are not present in the original light signal.
In applications sensitive to long-range time correlations, one could minimize this effect by preferably mounting the SPAD chip in a good contact with a large thermal mass (this however can only be done by the SPAD manufacturer) or by limiting the detection frequency or by using a SPAD with a lower operating voltage and therefore lower heat production.

\section*{Robustness test}

It is a well known fact that a typical photomultiplier (operated at a nominal high voltage) becomes irreversibly destroyed when exposed to daylight. Even when disconnected from the high voltage, an exposure to a normal daylight will cause a significant (albeit in that case reversible) rise of the noise level which can be cured only after hours of operation in dark. This is to be contrasted to SPADs that are immune to a strong light due to inherent limit of the sustainable avalanche current.
\\
 
In order to test the immunity of the SPCM detector against exposure to light, a fully operational detector (all supply voltages at nominal level) has been repeatedly exposed to strong daylight (direct hit of the Sun light to the SPAD) for prolonged periods of time ($>$10 minutes) and then suddenly brought to normal testing conditions. During the strong incident light, the detector was temporarily paralyzed (no counts at the output). However, not only that the detector was not destroyed by the strong light, but we could not find any evidence of change of its characteristics in subsequent tests. Notably, the noise level in darkness (dark counts rate) and the detection efficiency were restored to nominal level only seconds after removal of the overwhelming illumination. 

\section*{Conclusion}

A photon detector comprising novel active quenching circuit and the well known SPAD C30902 has been built and tested. Its main characteristics are: short dead time ($<$40ns), high counting capability, low noise, good detection efficiency, immunity against strong light and general physical robustness. The active quenching circuit can be built from easily obtainable, low-cost parts and requires only a simple adjusting procedure described in this paper. The main drawback of the present design is its inherent limit of operating the SPAD at an overvoltage less than 5V above the breakdown. With the popular SPAD C30902 from PerkinElmer this sets an absolute limit on achievable photon detection efficiency to approximately 25\% at the peak sensitivity wavelength (830nm). Because of its good overall performance and simplicity we believe that this detector will find its use in various applications.\\

This work was supported by Ministry of science education and sports of Republic of Croatia, contract number  098-0352851-2873 .

\end{document}